\documentclass{baltzer}
\usepackage{epsfig}
\begin{document}
\begin{frontmatter}
%
\title{Producing Slow Antihydrogen for a Test of CPT Symmetry
 with ATHENA}
\author[H]{M.C. Fujiwara,\thanks{e-mail:Makoto.Fujiwara@cern.ch}}
\author[D]{M. Amoretti,}
\author[J]{C. Amsler,}
\author[F]{G. Bendiscioli,}
\author[B]{G. Bonomi,}
\author[C]{A. Bouchta,}
\author[A]{P. Bowe,}
\author[D]{C. Carraro,}
\author[I]{M. Charlton,}
\author[I]{M. Collier,}
\author[C]{M. Doser,}
\author[F]{V. Filippini,}
\author[C]{K. Fine,}
\author[F]{A. Fontana,}
\author[H]{R. Funakoshi,}
\author[F]{P. Genova,}
\author[J]{D. Gr\"{o}gler,}
\author[A]{J.S. Hangst,}
\author[H]{R.S. Hayano,}
\author[H]{H. Higaki,}
\author[E]{M.H. Holzscheiter,}
\author[D]{W. Joffrain,}
\author[I]{L. Jorgensen,}
\author[D]{V. Lagomarsino,}
\author[C]{R. Landua,}
\author[G]{C. Lenz Cesar,}
\author[J]{D. Lindel\"{o}f,}
\author[B]{E. Lodi-Rizzini,}
\author[D]{M. Macri,}
\author[J]{N. Madsen,}
\author[D]{G. Manuzio,}
\author[C]{M. Marchesotti,}
\author[F]{P. Montagna,}
\author[J]{H. Pruys,}
\author[J]{C. Regenfus,}
\author[C]{P. Riedler,}
\author[F]{A. Rotondi,}
\author[C]{G. Rouleau,}
\author[F]{P. Salvini,}
\author[D]{G. Testera,}
\author[I]{D.P. van der Werf,}
\author[D]{A. Variola,}
\author[B]{L. Venturelli,}
\author[I]{T. Watson,}
\author[H]{T. Yamazaki,}
\author[H]{and Y. Yamazaki}

(ATHENA collaboration)

\address[A]{University of Aarhus, Denmark,}
\address[B]{Brescia University \& INFN, Italy,}
\address[C]{CERN, Switzerland,}
\address[D]{Genoa University \& INFN,}
\address[E]{Los Alamos Nat. Lab., USA,}
\address[F]{Pavia University \& INFN, Italy}
\address[G]{Fed. Univ. Rio de Janeiro (UFRJ), Brasil,}
\address[H]{University of Tokyo, Japan,}
\address[I]{University of Wales Swansea, UK,}
\address[J]{University of Zurich, Switzerland}
\runningauthor{M.C. Fujiwara {\it et al.}} \runningtitle{Producing
Slow Antihydrogen for a Test of CPT}
%
\begin{abstract}  
The ATHENA experiment at the Antiproton Decelerator facility at
CERN aims at testing CPT symmetry with antihydrogen. An overview
of the experiment, together with preliminary results of
development towards the production of slow antihydrogen are
reported.

\end{abstract}
\begin{keywords}  
Antiproton, Positron, Antihydrogen, Penning trap, CPT
\end{keywords}
\classification{} 
\end{frontmatter}
\section{Introduction}

Testing symmetries is one of the important subjects in physics.
While P (parity) and CP (charge-parity) are known to be violated,
CPT (charge-parity-time reversal) is believed to be conserved by
virtue of the CPT theorem~\cite{cpt}. The assumptions of the CPT
theorem do not, however, apply to some extensions of the Standard
Model including, notably, string theories~\cite{kostel99}. The
possibility of large extra dimensions~\cite{extra} may even lead
to CPT violation at the energy scale much lower than the Planck
scale. Several accurate tests of CPT invariance have been
performed on leptons and hadrons, and in exotic atoms~\cite{PDG},
yet given its fundamental importance, CPT should be tested in all
particle sectors.

The first production of antihydrogen, a bound system of antiproton
($\overline{p}$) and positron ($e^+$), was reported at
CERN~\cite{bauer96}, and later at Fermilab~\cite{fermi98}. These
anti-atoms, created at high velocity, annihilated almost
immediately after their production, leaving little time to study
their properties. It is the goal of ATHENA (AnTiHydrogEN
Apparatus) to produce a large quantity of slow antihydrogen to
study its properties and, via comparison with its well-studied
matter counterpart, to make precision tests of the CPT and other
symmetries of nature~\cite{MHreview}. Phase 1 of ATHENA focuses on
production and identification of slow antihydrogen. In this paper,
we describe the overview of the experiment together with recent
progress towards making slow antihydrogen.

\section{ATHENA Overview}
ATHENA is one of the three experiments at CERN's newly constructed
Antiproton Decelerator (AD) facility. ASACUSA~\cite{asacusa} and
ATRAP~\cite{atrap} are other experiments studying antiprotonic
atom spectroscopy/collision, and antihydrogen, respectively. The
AD, commissioned for physics in July, 2000, provides a pulse of a
few $\times 10^7$ $\overline p$ at 100 MeV/c (5.3 MeV in kinetic
energy) roughly every two minutes, with its performance steadily
improving.

\begin{figure}[h]
\begin{center}
\mbox{\epsfig{file=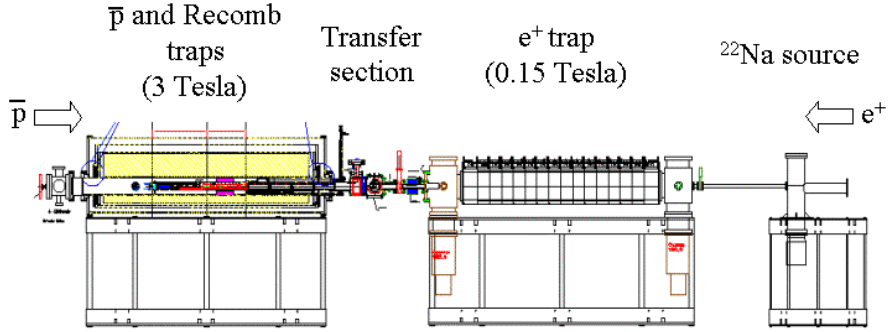,height=4.5cm}}
\end{center}
\caption{The ATHENA Apparatus}
\end{figure}

The ATHENA experiment consists of four major components: a
catching trap, a recombination trap, a positron accumulator, and
the detection system (Fig.~1). Antiprotons, first trapped in the
catching trap and cooled by electrons, are subsequently moved to
the recombination trap. Positrons are separately trapped in the
positron accumulator, and then transferred to the recombination
trap, where they are merged with antiprotons. Antihydrogen, formed
via recombination of two ingredients, escapes the trap confinement
and collides with the wall, thereby annihilating. Antihydrogen
annihilation will be detected by the ATHENA vertex detector,
consisting of Si micro-strips and segmented CsI crystals.
In addition, several other detectors are used to study the
trapping and cooling of antiprotons, which include a segmented Si
beam counter, a Faraday cup, a CCD camera, and external
scintillators coupled either to photo-multipliers or HPDs (hybrid
photodiodes)~\cite{hpd}.

\section{Antiproton Catching and Cooling}

Antiprotons are trapped in an open end-cap Penning
trap~\cite{pbartrap} in which the radial motion is confined by a 3
Tesla magnetic field in a superconducting solenoid, and the axial
motion, by an electrostatic potential ranging from 5 to 10 kV. A
pulsed beam of 5.3 MeV antiprotons from the AD passes through a
segmented Si beam detector, and slowed by a stainless steel vacuum
window and an Al degrader to less than 10 keV upon injection into
the catching trap. The entrance side of the trap potential is
opened for a few hundred ns when the beam arrives, in order to
allow the antiprotons to enter the trap.

\begin{figure}[h]
\begin{center}
\mbox{\epsfig{file=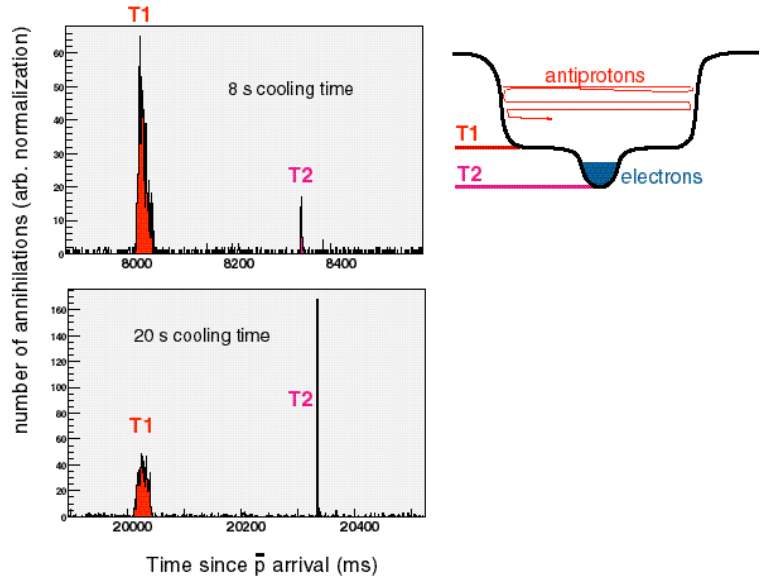,width=10cm}}
\end{center}
\caption{Electron cooling of antiprotons. }
\end{figure}

\begin{figure}[h]
\begin{center}
\mbox{\epsfig{file=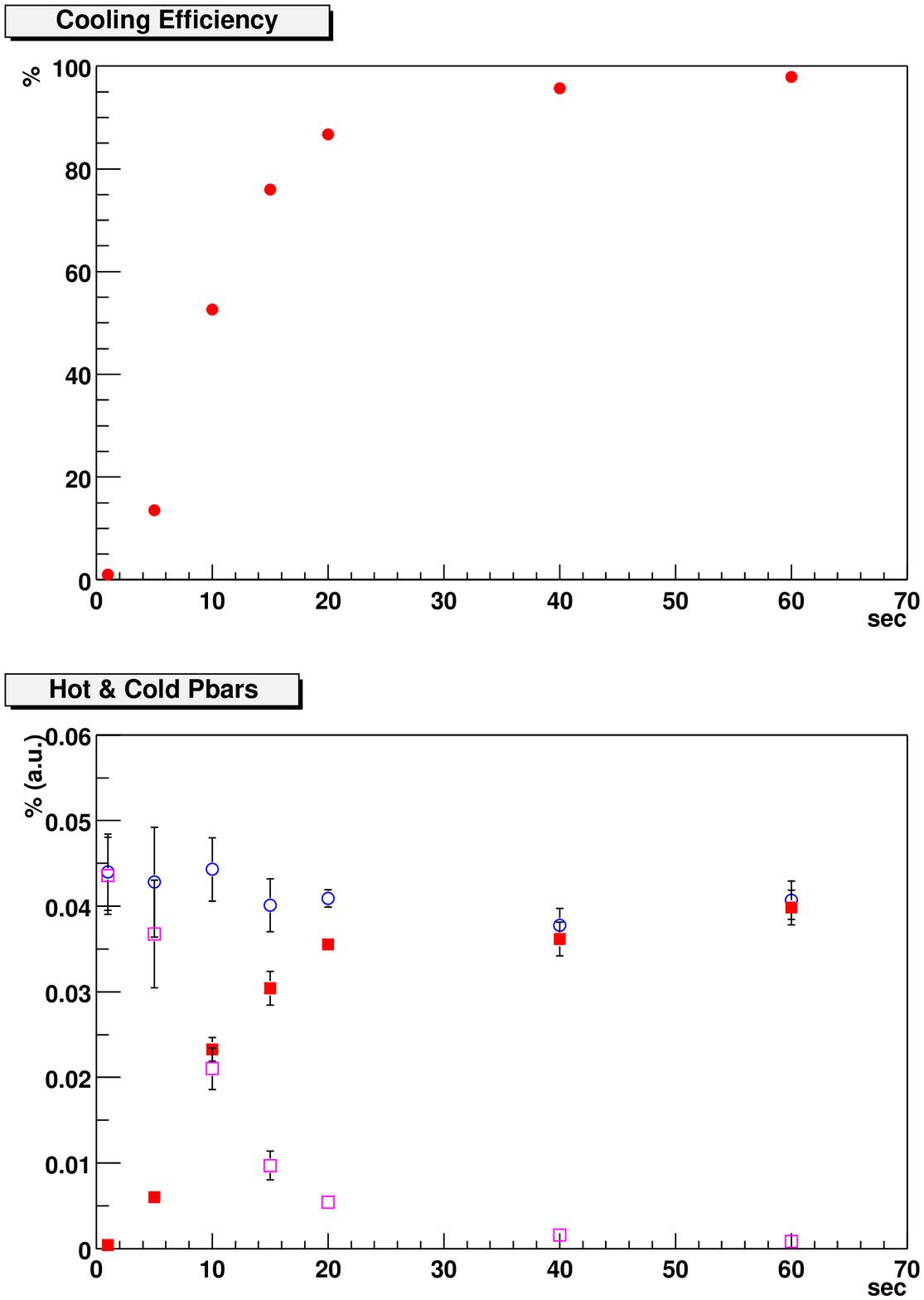,width=9cm}}
\end{center}
\caption{Top: The ratio of antiprotons cold/(cold+hot) as a
function of cooling time. Bottom: The numbers of cold (plotted
with filled square), hot (open square), and cold+hot (open circle)
antiprotons, normalized to the beam intensity.}
\end{figure}

A few $\times 10^8$ electrons, preloaded in the central part of
the trap, interact with trapped antiprotons and cool them to
$\sim$eV energy or below. Figure 2 illustrates the demonstration
of electron cooling in the ATHENA catching trap. The outer
potential wall (5 kV) and the inner potential wall (40 V) were
opened at time T1 and T2, respectively, dumping the antiprotons
onto a foil. The annihilations of the released antiprotons at each
time, detected by the scintillators, indicate the number of
uncooled (``hot''), and cooled (``cold'') antiprotons. A
systematic measurement of the cooling process is shown in Fig.~3.
The efficiency of electron cooling depends rather sensitively on
the electron loading conditions, an effect which requires further
investigation. The lifetime of trapped antiprotons is longer than
several hours: it is possible to store them in our trap over
night, if necessary.

\section{Detecting Antihydrogen}

The formation of antihydrogen will take place in the recombination
trap, to which cooled antiprotons as well as positrons, the latter
trapped with the nitrogen buffer gas method~\cite{surko}, are
transferred. Several recombination schemes are discussed in
Ref.~\cite{MHreview}.

\begin{figure}[t]
\begin{center}
\mbox{\epsfig{file=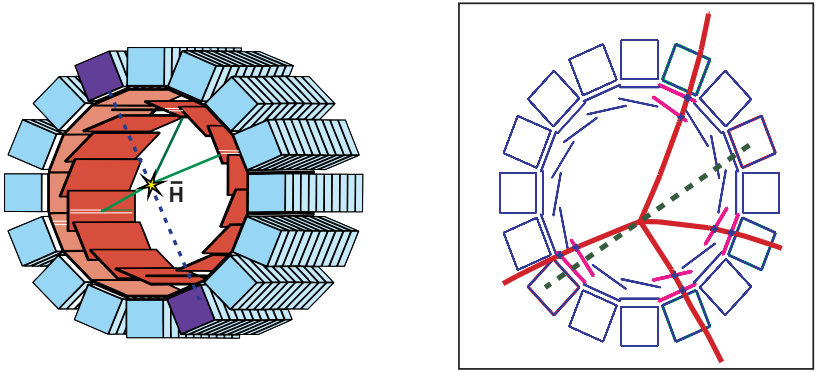,width=0.8\textwidth}}
\end{center}
\caption{Left: Schematic view of the ATHENA vertex detector.
Right: A GEANT simulated antihydrogen event showing  511 keV
$\gamma$s (dashed line) and charged pions (solid line).
\label{fig:golden}}
\end{figure}

The ATHENA vertex detector surrounds the recombination region. The
production of antihydrogen will be identified by the simultaneous
detection (within $\sim 2 \ \mu s$) of charged pions from the
$\overline p$ annihilation, and back-to-back 511 keV $\gamma$ rays
from the $e^+$ annihilation, both originating from the same vertex
(Fig.~\ref{fig:golden}). Two layers of double sided Si
micro-strips, and 16 $\times$ 12 pure CsI crystals coupled to
photodiodes, a total of 8192 channels of signals, are read out,
and the events will be reconstructed off-line.

Figure~\ref{fig:golden} (right) is an example of simulated
antihydrogen events, shownig the detector front view of four
charged pion tracks and back-to-back $\gamma$s. A dominant
source of 511 keV background is expected to come from the
annihilation of $e^+$ produced in $\pi ^0$ decay and
subsequent electromagnetic showers in the surrounding
material.
These $\gamma$s, however, would have angular correlations that are
generally random, hence can be discriminated against by plotting
the opening angles of two $\gamma$s with respect to the
reconstructed charged vertex. Hence, a high degree of CsI
segmentation is important in overcoming this background.

\begin{figure}[b]
\begin{center}
\mbox{\epsfig{file=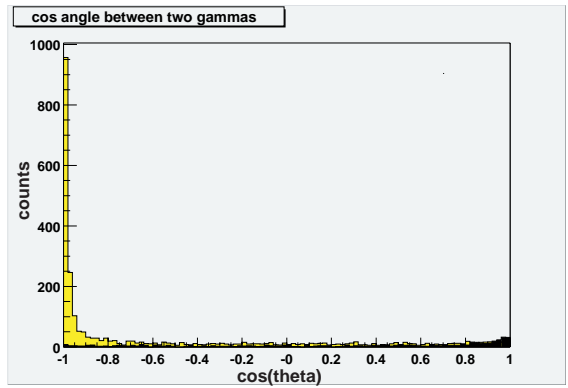,width=0.55\textwidth}}
\end{center}
\caption{GEANT simulation of the opening angle $\theta$ between
two 511 keV $\gamma$s with respect to the reconstructed charged
vertex, for the antihydrogen events (light histogram) and the
random background (dark histogram).\label{fig:cos}}
\end{figure}

Preliminary simulations have been performed, assuming a background
of simultaneous $\bar{p}$ and $e^+$ annihilations with the same
frequency as antihydrogen, but occurring randomly in different
points on the trap walls. The results indicate that a
signal-to-background ratio greater than $70$ may be obtained for
the opening angle $\theta$ of $-1 \le \cos \theta \le - 0.9 $ (see
Fig.~\ref{fig:cos}~\cite{rotondi}).

\section{Summary and Outlook}

Since the physics start in July 2000, the ATHENA experiment has
succeeded in trapping and cooling more than 10$^4$ antiprotons per
AD pulse, accumulating more than 10$^7$ positrons in one minute,
and transferring them to the recombination trap (although not yet
at the same time). In year 2001, we will make our first attempt to
merge the two species to produce slow atoms of antihydrogen.


%

\begin{thebibliography}{00} 

\bibitem{cpt}
G. Luders, Ann. Phys. {\bf 2}, 1 (1957).

\bibitem{kostel99}
see, e.g. {\it CPT and Lorentz Symmetry} ed. V. A. Kosteleck\'{y},
World Scientific, Singapore, 1999.

\bibitem{extra}
N. Arkani-Hamed, S. Dimopoulos and G. Dvali, Phys. Lett. B {\bf
429}, 263 (1998).

\bibitem{PDG}
Review of Particle Physics, D. E. Groom {\it et al.}, Eur. Phys.
Jour.C {\bf 15}, 1 (2000).

\bibitem{bauer96}
G. Bauer {\it et al.,} Phys. Lett. B {\bf 368}, 251 (1996).

\bibitem{fermi98}
G. Blanford {\it et al.,} Phys. Rev. Lett. {\bf 80}, 3037 (1998).

\bibitem{MHreview}
M. H. Holzscheiter and M. Charlton, Rep. Prog. Phys. {\bf 62}, 1
(1999).

\bibitem{asacusa}
CERN Proposal SPSC 97-19/P307; see also Y. Yamazaki, and E.
Widmann, this volume.

\bibitem{atrap}
CERN Proposal SPSC 97-8/P306.

\bibitem{hpd}
M.C. Fujiwara, D. Grassi and M. Marchesotii, to be published in
Nucl. Instrum. and Meth.

\bibitem{pbartrap} G. Gabrielse {\it et al.,} Phys. Rev. Lett.
{\bf 57}, 2504 (1986); {\bf 63}, 1360 (1989);\\ M. H. Holzscheiter
{\it et al.,} Phys. Lett. A {\bf 2214}, 279 (1996).

\bibitem{surko}
T. J. Murphy and C. M. Surko, Phys. Rev. A {\bf 46} 5696 (1992).

\bibitem{rotondi}
A. Rotondi {\it et al.}, ATHENA technical report (unpublished).

\end{thebibliography}
\end{document}